\documentclass[%
 aip,
 amsmath,amssymb,
 reprint,%
]{revtex4-1}

\usepackage{graphicx}
\usepackage{dcolumn}
\usepackage{bm}

\usepackage[utf8]{inputenc}
\usepackage[T1]{fontenc}
\usepackage{mathptmx}
\usepackage{etoolbox}
\usepackage{subcaption}

\makeatletter
\def\@email#1#2{%
 \endgroup
 \patchcmd{\titleblock@produce}
  {\frontmatter@RRAPformat}
  {\frontmatter@RRAPformat{\produce@RRAP{*#1\href{mailto:#2}{#2}}}\frontmatter@RRAPformat}
  {}{}
}%
\makeatother
\begin{document}

\title{Inhalation--exhalation asymmetry in pressure-driven airflow through a population-derived median unilateral human nasal cavity}
\author{Makesh Singh}
\author{Shashwat Bhattacharya}
\author{Gaurav Bhutani}%
 \email{gaurav@iitmandi.ac.in}
\affiliation{%
School of Mechanical and Materials Engineering\\
Indian Institute of Technology Mandi\\
Mandi 175075, Himachal Pradesh, India
}%

\begin{abstract}
\noindent
\textbf{ABSTRACT}\\
This study investigates inhalation--exhalation asymmetry in population-derived median unilateral human nasal cavity under pressure-driven breathing using time-resolved computational fluid dynamics (CFD). A standardized geometry reconstructed from averaged CT data of 30 adults was subjected to a sinusoidal nasopharyngeal pressure waveform over a complete breathing cycle. Low, intermediate, and high pressure amplitudes were considered to quantify breathing-phase-dependent changes in flow rate, Reynolds number, wall shear stress, nasal resistance, and olfactory-region airflow. For identical imposed pressure magnitudes, inhalation produced a consistently larger flow rate and lower aerodynamic resistance than exhalation, demonstrating a direction-dependent pressure--flow response. Increasing breathing intensity increased Reynolds number and nasal resistance while largely preserving the spatial pattern of the flow. During inhalation, stronger airflow penetrated further into the superior nasal cavity and produced higher and more localized wall shear stress, whereas exhalation exhibited smoother airflow redistribution. The breathing-phase asymmetry also affected regional ventilation: the fraction of total airflow reaching the olfactory region increased with inspiratory intensity and reached approximately 6\% under the high-pressure condition, while the expiratory fraction remained comparatively insensitive to breathing intensity. These results demonstrate that pressure-driven transient simulations naturally capture geometry-dependent inhalation--exhalation differences that are suppressed when identical flow rates are prescribed for the two breathing phases, with consequences for both global nasal resistance and regional airflow delivery.
\end{abstract}

\maketitle
\section{Introduction}
Nasal breathing warms and humidifies inspired air, filters inhaled particulates, and transports odorants to the olfactory epithelium before air enters the lower airways \citep{subramaniam1998computational,Eccles2000}. 
These functions arise from the complex three-dimensional anatomy of the nasal cavity, particularly the nasal valve, turbinates, and olfactory cleft, which produce highly non-uniform velocity, pressure, and wall shear stress distributions \citep{doorly2008mechanics,zhao2014normal}. The intricate geometry, together with the cyclic nature of respiration, makes experimental characterization of nasal airflow challenging, and computational fluid dynamics (CFD) has therefore become an important tool for investigating nasal aerodynamics in both healthy and diseased airways.

CFD studies have provided detailed insight into airflow organization, pressure loss, nasal resistance, particle transport, and olfactory ventilation using anatomically realistic nasal geometries reconstructed from medical imaging \citep{subramaniam1998computational,kim2013patient}. Subsequent investigations extended these analyses to patient-specific anatomies, statistical shape models, and pathological conditions, demonstrating the strong influence of geometry on airflow distribution and regional ventilation \citep{keustermans2018high,basu2018computational,asama2022analysis}. Although steady-state simulations remain widely used, transient CFD has shown that cyclic breathing produces breathing-phase-dependent variations in airflow, pressure distribution, and wall shear stress that cannot be fully captured under steady assumptions \citep{lee2010unsteady,gu2019,moniripiri2021numerical, shahed2024analysis}.

Most numerical studies consider bilateral nasal cavities, in which airflow is partitioned between two parallel passages. In healthy adults, the nasal cycle produces alternating congestion and decongestion that modifies the resistance of each side while maintaining relatively stable overall nasal resistance \citep{Eccles2000,zhao2014normal}. However, airflow may predominantly pass through a single nasal cavity during the nasal cycle or in pathological conditions such as septal deviation, partial obstruction, and nasal polyps \citep{patel2015simulating,frank2019computational}. Under these conditions, the loss of the parallel flow pathway increases unilateral resistance and alters regional airflow organization. Unilateral models therefore provide a useful framework for isolating side-specific aerodynamic behaviour and systematically investigating local resistance and airflow distribution.

Most CFD studies prescribe the inlet flow rate, whereas physiological breathing is driven by pressure differences generated by the respiratory system. Pressure-driven formulations allow the flow rate to develop naturally from the interaction between the imposed pressure difference and the nasal geometry, providing a more physiological representation of breathing \citep{faramarzi2014numerical,gu2019,shahed2024analysis}.
Despite extensive CFD investigations, systematic transient studies of representative unilateral nasal geometries under pressure-driven breathing remain limited. Many computational investigations employ subject-specific anatomies, for which the predicted airflow characteristics may be strongly influenced by individual anatomical features. Population-derived standardized geometries provide a complementary framework by representing central anatomical tendencies across multiple subjects, allowing fundamental aerodynamic mechanisms to be examined without attributing the results to the anatomy of a particular patient.
Such models are particularly useful for investigating directional effects arising from the intrinsic geometry of the nasal passage.

The present study therefore employs the population-derived standardized median nasal model of~\citet{liu2009standardized}, constructed from CT measurements of 30 adults with normal nasal anatomy, to investigate pressure-driven airflow through a representative unilateral nasal passage over a complete respiratory cycle.
Three pressure amplitudes representing low, intermediate, and high breathing effort are considered. 
By prescribing pressure rather than flow rate, the inhalation and exhalation flow rates are allowed to emerge from the interaction between the imposed pressure difference and nasal geometry. The analysis focuses on whether this formulation produces an inherent directional asymmetry in flow rate and resistance and examines its consequences for Reynolds number, wall shear stress, pressure loss, and airflow delivery to the olfactory region.

The outline of the paper is as follows. Section~\ref{sec:Model} describes the flow geometry, computational mesh, and the numerical methods employed in our simulations.  In Sec.~\ref{sec:Results}, we analyse the simulation results, focusing on the breathing dynamics, nasal resistance, and airflow in the olfactory region. We discuss the implications of our results in Sec.~\ref{sec:Discussion}, and conclude in Sec.~\ref{sec4}.

\maketitle
\section{Model description} \label{sec:Model}
\subsection{Geometry and computational mesh}

The computational geometry was not derived from an individual patient. It was reconstructed from the standardized median nasal model reported by \citet{liu2009standardized}, which was developed from CT measurements of 30 adults (13 male and 17 female; age range 17--78 years, median 45 years) with normal nasal anatomy. The CT examinations had been acquired during clinical work-up for migraine, cluster, or tension headaches, conditions not associated with accepted anatomical abnormalities of the nasal cavity; subjects with obstructing lesions or septal deviation were excluded from the standardized model. Thus, the geometry represents population-derived median unilateral nasal anatomy rather than the anatomy of a single subject.

\begin{figure*}[htbp]
	\centering
	\includegraphics[width=0.9\linewidth]{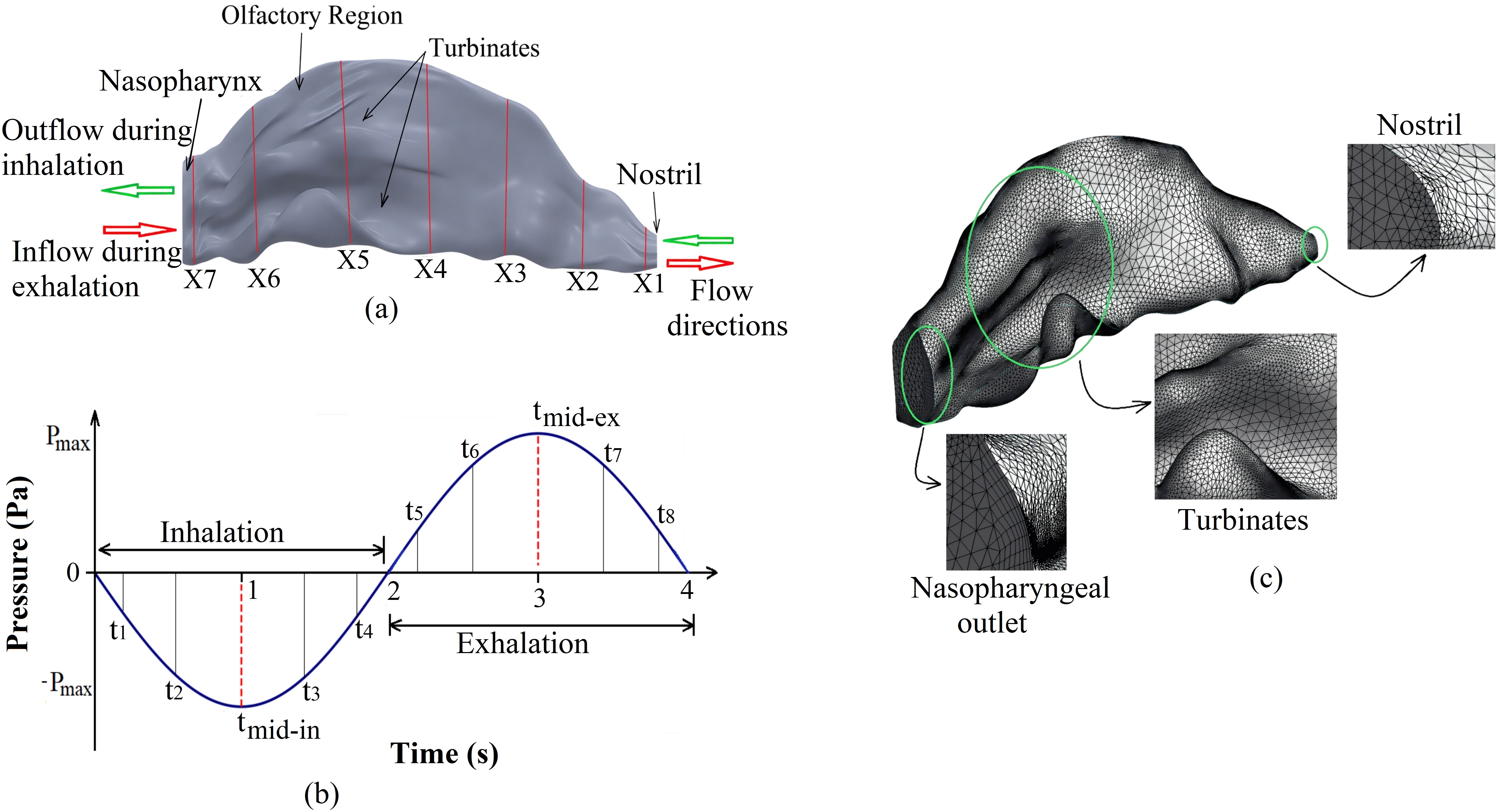}
	\caption{Right unilateral nasal cavity used in the simulations: (a) major anatomical regions and airflow directions during inhalation and exhalation; (b) coronal sections X1--X7 used for regional analysis and (c) regional meshing.}
	\label{fig:uncreconstruction}
\end{figure*}

The averaged coronal profiles were digitized in SolidWorks\textsuperscript{\textregistered} to generate a smooth, watertight model of the right nasal passage. The nasal valve, turbinates, olfactory region, and nasopharyngeal outlet were retained, while small-scale surface irregularities were omitted. The resulting cavity was approximately 60~mm long and 35~mm wide. Seven coronal sections (X1--X7) were defined for regional analysis, as shown in Fig.~\ref{fig:uncreconstruction}.

To characterize the flow regime, the Reynolds number was evaluated at the seven coronal sections as:
\begin{equation}
\mathrm{Re}=\frac{U_{\mathrm{avg}}D_h}{\nu},
\label{eq:reynolds}
\end{equation}
where $U_{\mathrm{avg}}$ is the area-averaged velocity across each section, $D_h$ is the corresponding hydraulic diameter, and $\nu$ is the kinematic viscosity of air.

The computational domain was discretized using an unstructured tetrahedral mesh generated in ANSYS\textsuperscript{\textregistered} Meshing. Local refinement was applied near the nostril, nasal valve, turbinates, cavity walls, and nasopharyngeal outlet to resolve regions of strong velocity gradients and accurately evaluate pressure loss and wall shear stress (Fig.~\ref{fig:uncreconstruction}(c)).

\subsection{Governing equations and numerical method}

Airflow was modelled as three-dimensional, incompressible, laminar, and Newtonian. The governing continuity and momentum equations are
\begin{equation}
	\nabla \cdot \mathbf{u}=0,
	\label{eq:continuity}
\end{equation}
and
\begin{equation}
	\rho\left(
	\frac{\partial \mathbf{u}}{\partial t}
	+\mathbf{u}\cdot\nabla\mathbf{u}
	\right)
	=-\nabla p+\mu\nabla^{2}\mathbf{u},
	\label{eq:momentum}
\end{equation}
where $\mathbf{u}$ is the velocity, $p$ is the pressure, and $\rho$ and $\mu$ are the density and dynamic viscosity of air, respectively.

Under resting breathing conditions, nasal airflow is generally laminar or transitional, although localized transitional effects may arise near constrictions such as the nasal valve and in downstream regions. Previous studies have shown that laminar models provide pressure-drop and velocity predictions comparable to higher-fidelity approaches for the principal nasal airflow characteristics considered here \citep{li2017computational,schillaci2022importance}. Unsteady laminar simulations have also been shown to capture breathing-phase-dependent airflow redistribution and wall shear stress during normal breathing \citep{lee2010unsteady,shahed2024analysis}. A laminar formulation was therefore adopted for all breathing conditions.

The equations were solved in ANSYS\textsuperscript{\textregistered} Fluent using the finite-volume method. Convective terms were discretized using a second-order upwind scheme, diffusion terms using central differencing, and a second-order time implicit scheme was employed overall. Pressure--velocity coupling was achieved using the SIMPLE algorithm with second-order pressure interpolation. Iterations within each time step were continued until the continuity and momentum residuals fell below $10^{-4}$.

\subsection{Boundary and operating conditions}

A sinusoidal gauge-pressure waveform (Fig.~\ref{fig:uncreconstruction}(b)) was prescribed at the nasopharyngeal boundary using a user-defined function:
\begin{equation}
	P(t)=
	\begin{cases}
		P_{\mathrm{in}}\,
		\sin\!\left(\dfrac{\pi t}{T_{\mathrm{in}}}\right),
		& 0\leq t<T_{\mathrm{in}}, \\[8pt]
		P_{\mathrm{ex}}\,
		\sin\!\left(\dfrac{\pi(t-T_{\mathrm{in}})}
		{T_{\mathrm{ex}}}\right),
		& T_{\mathrm{in}}\leq t<T_{\mathrm{in}}+T_{\mathrm{ex}},
	\end{cases}
	\label{eq:pressure_waveform}
\end{equation}
where $P_{\mathrm{in}}<0$ and $P_{\mathrm{ex}}>0$ are the peak inhalation and exhalation pressures, respectively, and $T_{\mathrm{in}}$ and $T_{\mathrm{ex}}$ are the corresponding breathing-phase durations. The breathing period was
$T=T_{\mathrm{in}}+T_{\mathrm{ex}}=4~\mathrm{s}$, with an inspiration-to-expiration ratio of $1{:}1$. 
Three pressure amplitudes representing low, intermediate, and high breathing effort were investigated, with equal inhalation and exhalation pressure magnitudes, $|P_{\mathrm{in}}|=|P_{\mathrm{ex}}|=$ 18, 60, and 130~Pa, respectively.

Zero gauge pressure was imposed at the nostril. The airway walls were assumed rigid and stationary, with a no-slip boundary condition. Air density and dynamic viscosity were set to
$\rho=1.2041~\mathrm{kg\,m^{-3}}$ and
$\mu=1.81\times10^{-5}~\mathrm{Pa\,s}$, respectively.

\subsection{Performance metrics}

To ensure consistency with clinical rhinomanometry practice, unilateral nasal resistance was evaluated from the \emph{inhalation-phase averages} of the pressure drop (across the cavity) and volumetric flow rate as:
\begin{equation}
R=
\frac{\left\langle \Delta P\right\rangle_{\mathrm{in}}}
{\left\langle Q\right\rangle_{\mathrm{in}}},
\label{eq:resistance}
\end{equation}
where $\left\langle \Delta P\right\rangle_{\mathrm{in}}$ and
$\left\langle Q\right\rangle_{\mathrm{in}}$ denote the corresponding averages over the inhalation phase.




\subsection{Numerical sensitivity analysis}

Grid- and time-step-independence studies were performed for the high-pressure breathing condition, which produced the largest velocities and therefore represented the most stringent numerical case.

\begin{figure}[htbp]
	\centering
	\includegraphics[width=1\linewidth]{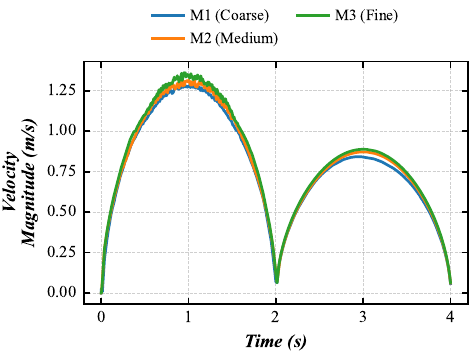}
	\caption{Area-averaged velocity at the nasopharynx for meshes M1--M3.}
	\label{fig:velocitygridindsinglecolumn}
\end{figure}

Three unstructured tetrahedral meshes, M1--M3, were considered (Table~\ref{tab:mesh}). The area-averaged velocity histories at the nasopharynx are compared in Fig.~\ref{fig:velocitygridindsinglecolumn}. Refinement from M1 to M2 changed the peak velocity by approximately $2.9\%$, whereas refinement from M2 to M3 changed it by only $0.8\%$. The M2 mesh was therefore selected for the remaining simulations.

\begin{table}[htbp]
	\centering
	\caption{Meshes used for the grid-independence analysis
($\Delta t = 10^{-4}\,\mathrm{s}$).}
	\label{tab:mesh}
	\begin{tabular}{ccccc}
		\hline
		Mesh &
		$N_e$ ($\times10^{6}$) &
		$N_n$ ($\times10^{6}$) &
		$\Delta$ (mm) &
		$\mathrm{CFL}_{\max}$ \\
		\hline
		M1 & 0.35 & 0.11 & 7 & 0.24 \\
		M2 & 0.57 & 0.16 & 4 & 0.43 \\
		M3 & 0.74 & 0.23 & 3 & 0.56 \\
		\hline
	\end{tabular}
\end{table}

Temporal sensitivity was subsequently evaluated on the M2 mesh using
$\Delta t=10^{-3}$, $10^{-4}$, and $10^{-5}~\mathrm{s}$, corresponding to 4000, 40\,000, and 400\,000 time steps per breathing cycle, respectively. Refining the time step from $10^{-3}$ to $10^{-4}~\mathrm{s}$ changed the peak velocity by approximately $3.5\%$, whereas further refinement to $10^{-5}~\mathrm{s}$ produced a change of only $0.7\%$. Consequently, the M2 mesh and $\Delta t=10^{-4}~\mathrm{s}$ were used for all simulations. The maximum CFL number remained below unity for all meshes.

\subsection{Validation}

\begin{figure}
	\centering
	\includegraphics[width=1\linewidth]{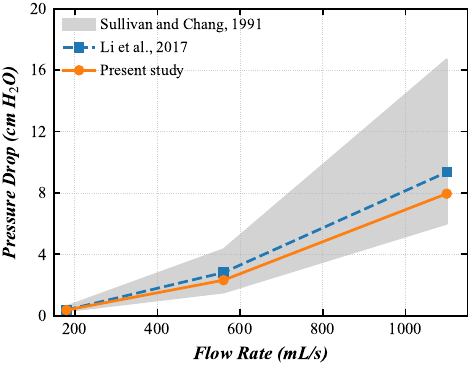}
	\caption{Validation of the predicted pressure-drop--flow relationship against the experimental measurements of \citet{sullivan1991steady} (gray shaded region) and the laminar CFD simulations of \citet{li2017computational} (blue dashed line).}
	\label{fig:validationgraph}
\end{figure}

The computational framework was validated by comparing the predicted pressure-drop--flow relationship with the experimental measurements of \citet{sullivan1991steady} and the laminar CFD simulations of \citet{li2017computational}, as shown in Fig.~\ref{fig:validationgraph}. For comparison with \citet{li2017computational}, additional steady-state simulations were performed using a constant-velocity inlet at the nostril and an outflow boundary condition at the nasopharynx, matching the reference boundary conditions. The predicted pressure drops lie within the experimentally reported range of \citet{sullivan1991steady} and agree closely with the numerical results of \citet{li2017computational} over the investigated flow rates. The close agreement with both experimental and numerical data demonstrates that the present computational framework accurately reproduces the pressure--flow characteristics of nasal airflow and further supports the suitability of the adopted laminar formulation for the breathing conditions considered.

\section{Results} \label{sec:Results}

The transient simulations are organized to examine how pressure-driven breathing produces inhalation--exhalation asymmetry and its consequences for (i) breathing dynamics, (ii) nasal resistance, and (iii) airflow delivery to the olfactory region.

\subsection{Breathing dynamics}

\begin{figure}[htbp]
	\centering
	\includegraphics[width=1\linewidth]{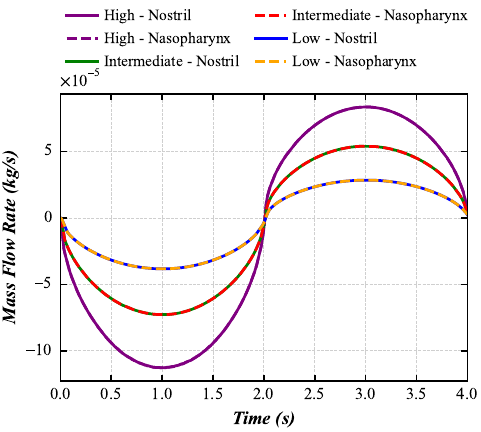}
	\caption{Transient mass flow rate at the nostril and nasopharynx during one breathing cycle for the three pressure-driven breathing conditions.}
	\label{fig:massflowrateinletoutlet}
\end{figure}

Figure~\ref{fig:massflowrateinletoutlet} shows the transient mass flow rate at the nostril and nasopharynx for the three breathing conditions.
The overlap between the nostril and nasopharynx mass flow rates confirms conservation of mass (what enters through nostril leaves through nasopharynx).
The flow varies periodically over the breathing cycle, with zero crossings corresponding to the transition between inhalation and exhalation. As expected, increasing the imposed pressure amplitude increases the peak mass flow rate while preserving the overall waveform.

A notable feature of the pressure-driven formulation is the asymmetry between inhalation and exhalation. 
For the same imposed pressure magnitude, inhalation produces a larger peak mass flow rate than exhalation, indicating direction-dependent resistance associated with the asymmetric nasal geometry. 
The present waveform assumes equal durations for the two breathing phases to isolate this aerodynamic response. Physiological breathing, however, generally has a shorter inhalation than exhalation phase \citep{motamedi2021reference}; hence, a larger inspiratory flow rate can still be consistent with equal inhaled and exhaled volumes over a complete breathing cycle.

\begin{figure*}[htbp]
	\centering
	\includegraphics[width=\linewidth]{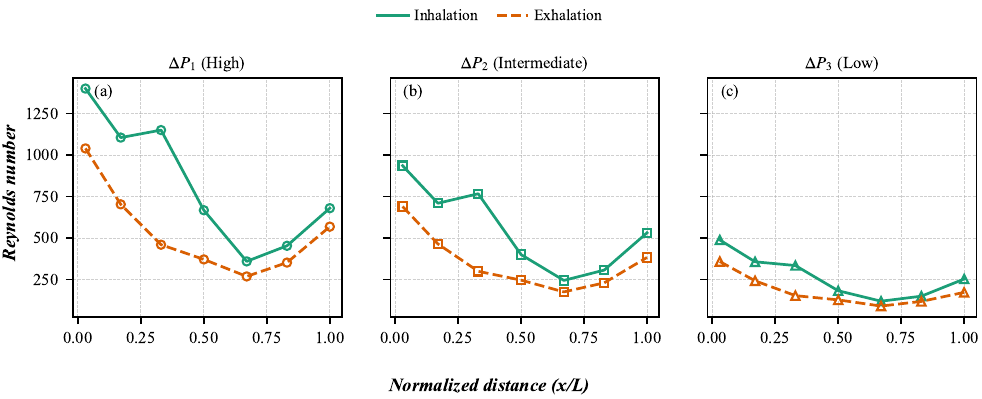}
	  \caption{Reynolds number distribution along the normalized nasal cavity length ($x/L$, measured from nose) during inhalation and exhalation for low-, intermediate-, and high-pressure breathing conditions.}
	\label{fig:reynoldnounccompare}
\end{figure*}

The corresponding Reynolds number distributions are shown in Fig.~\ref{fig:reynoldnounccompare}. For all breathing conditions, the Reynolds number decreases from the nasal valve toward the middle cavity before increasing again in the posterior region, reflecting the variation in local hydraulic diameter and flow acceleration/deceleration along the airway. Increasing the pressure amplitude primarily increases the magnitude of the Reynolds number while preserving its spatial distribution. Inhalation consistently produces higher Reynolds numbers than exhalation at corresponding locations because of the larger flow rates generated during the inspiratory phase, highlighting lower flow resistance during inhalation. Although Reynolds numbers increase under high-pressure breathing, the flow is expected to remain predominantly laminar, with localized transitional behaviour possible near geometric constrictions, consistent with previous PIV and DNS studies \cite{doorly2008mechanics}.

\begin{figure*}[htbp]
    \centering

    \begin{subfigure}{0.9\linewidth}
        \centering
        \includegraphics[width=\linewidth]{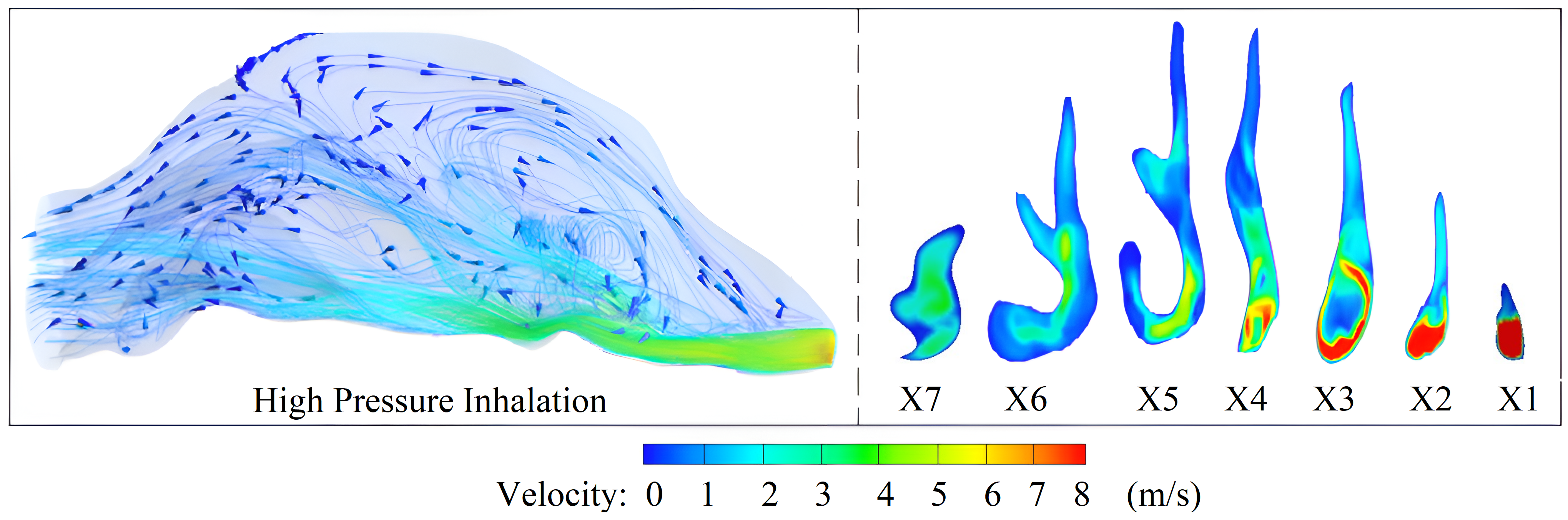}
        \caption{Mid-inhalation}
        \label{fig:streamlinesvelocitycontoursinhalation}
    \end{subfigure}

    \vspace{0.5em}

    \begin{subfigure}{0.9\linewidth}
        \centering
        \includegraphics[width=\linewidth]{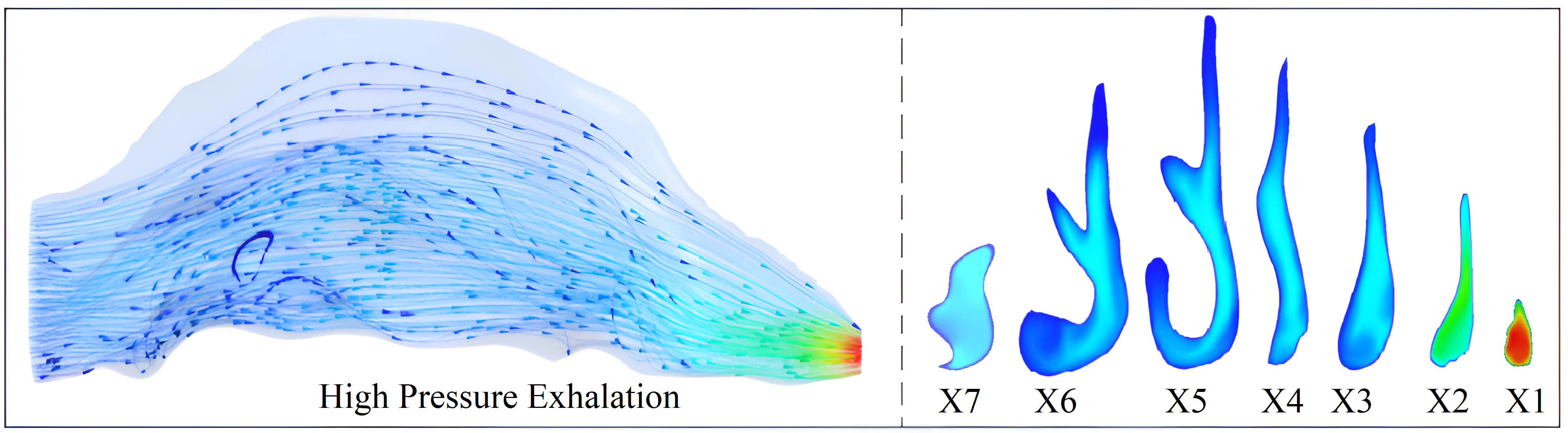}
        \caption{Mid-exhalation}
        \label{fig:streamlinesvelocitycontoursexhalation}
    \end{subfigure}

    \caption{Comparison of airflow during the high-pressure breathing condition at (a) mid-inhalation, and (b) mid-exhalation. Streamline patterns (left) and velocity-magnitude contours (right).}
    \label{fig:phase_airflow}
\end{figure*}

Since increasing breathing effort primarily amplifies the overall flow intensity, only the high-pressure condition is presented in Fig.~\ref{fig:phase_airflow} to illustrate breathing-phase-dependent airflow organization. During inhalation, air accelerates through the anterior nasal cavity and nasal valve before redistributing toward the middle and superior regions. 
The stronger inspiratory flow penetrates into the superior cavity, whereas exhalation produces a smoother and more diffuse redistribution from the nasopharynx toward the nostril. This breathing-phase-dependent organization suggests that inspiratory flow may promote greater airflow delivery to the olfactory region, as examined subsequently. This phase-dependent redistribution is consistent with literature~\citep{horschler2003numerical,han2023investigation} reporting enhanced secondary flow and higher in-plane velocity in the middle and superior meatus during the inhalation.

\subsection{Nasal resistance}

The average unilateral nasal resistance increased with breathing intensity, from $0.57~\mathrm{Pa{\cdot}s/mL}$ under the low-pressure condition to $1.00~\mathrm{Pa{\cdot}s/mL}$ under the intermediate-pressure condition and $1.22~\mathrm{Pa{\cdot}s/mL}$ under the high-pressure condition. These findings are consistent with ~\citet{van2021pressure}, who reported that nasal resistance increased with increasing flow rate and attributed the dominant pressure loss to geometric constrictions, particularly the nasal valve. These values were evaluated from the phase-averaged pressure drop and flow rate during inhalation, consistent with standard rhinomanometry practice, and lie within the reported healthy unilateral range of $0.36$--$1.36~\mathrm{Pa{\cdot}s/mL}$ \citep{eccles1996changes}.

\begin{figure}
    \centering
    \includegraphics[width=1\linewidth]{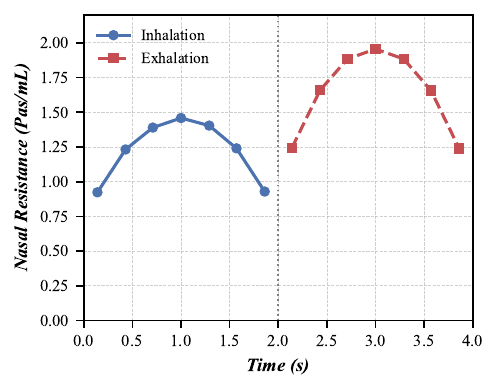}
    \caption{Temporal variation of nasal resistance during inhalation and exhalation under the high-pressure breathing condition.}
    \label{fig:nasal_resistance_time_instant}
\end{figure}

The temporal variation of nasal resistance for the high-pressure breathing condition is shown in Fig.~\ref{fig:nasal_resistance_time_instant}. During both phases, the resistance increases from the flow-reversal point, reaches a maximum near peak flow, and subsequently decreases. This variation reflects the nonlinear pressure--flow response of the nasal passage: as the flow rate increases, inertial pressure losses become progressively more important, producing a higher instantaneous resistance. The lower resistance during inhalation than exhalation is also consistent with the larger inspiratory flow rates obtained under identical pressure forcing, demonstrating the direction-dependent aerodynamic response of the asymmetric nasal geometry.

\begin{figure*}
    \centering
    \includegraphics[width=1\linewidth]{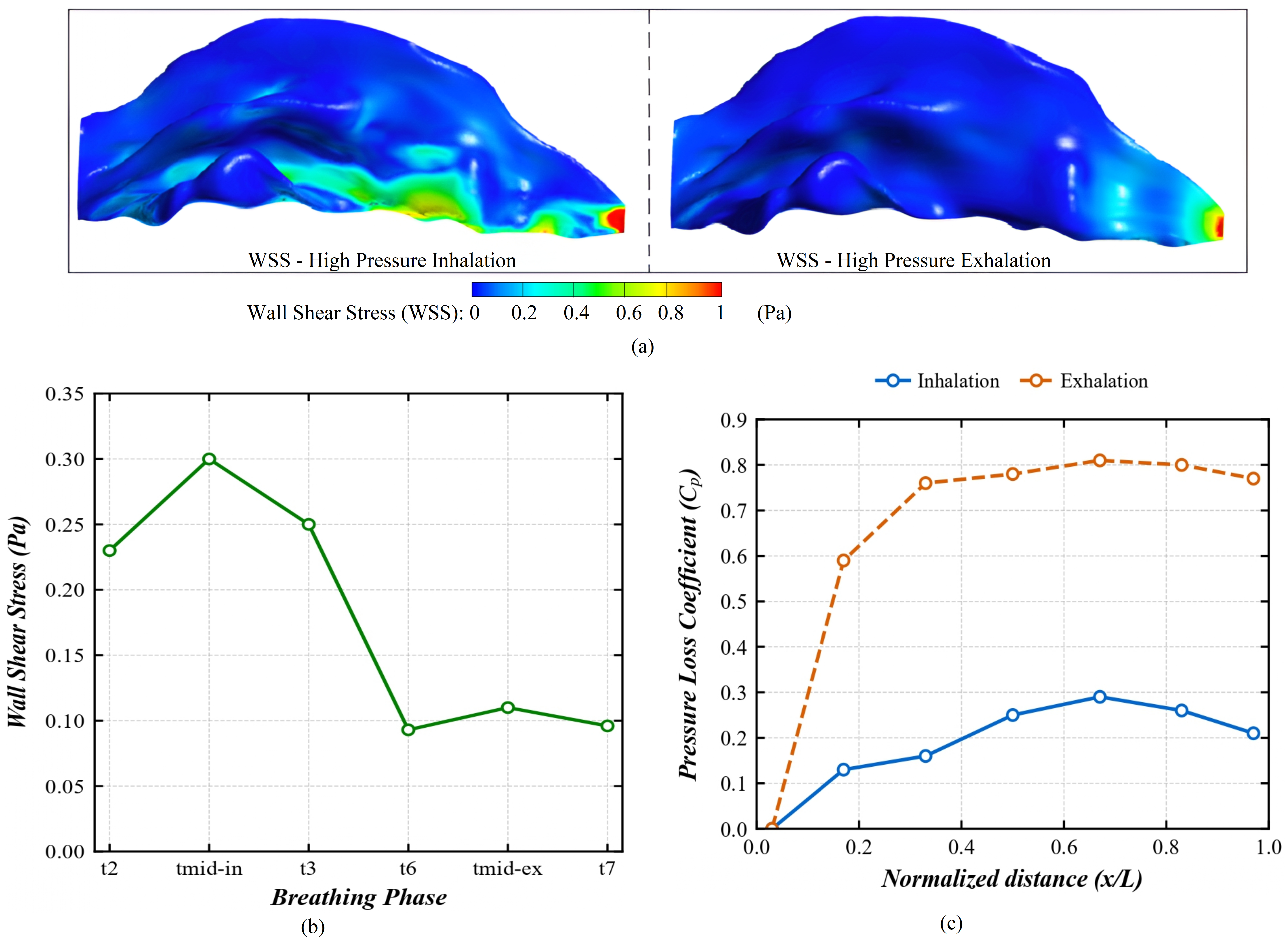}
    \caption{Contours of wall-shear stress for high-pressure (a) inhalation and (b) exhalation. (b) Wall shear stress versus the breathing phase, and (c) pressure loss coefficient ($C_p$) versus the normalized distance ($x/L$) for the high-pressure breathing condition.}
    \label{fig:wss_cp}
\end{figure*}

Further insight into the directional resistance is obtained from the high-pressure condition in Fig.~\ref{fig:wss_cp}. Wall shear stress (WSS) is concentrated near the nasal valve and turbinate surfaces and is higher during inhalation, indicating stronger local viscous loading (Fig.~\ref{fig:wss_cp}a, b). In contrast, the local pressure-loss coefficient is higher during exhalation over much of the nasal passage (Fig.~\ref{fig:wss_cp}c), consistent with the larger overall expiratory resistance. The opposing behaviour of WSS and pressure loss indicates that the inhalation--exhalation resistance asymmetry is not governed by wall viscous effects alone, but is strongly influenced by direction-dependent, geometry-induced losses associated with flow acceleration, turning, expansion, and redistribution through the nasal cavity. Higher wall shear stress was identified adjacent to the nasal valve and turbinate in earlier inhalation and exhalation studies~\citep{han2023investigation,tripathy2024effect}, whereas the nasal vestibule has also been shown to have high WSS~\citep{niegodajew2025flow}.

\subsection{Olfactory-region airflow}

\begin{figure*}
	\centering
	\includegraphics[width=1\linewidth]{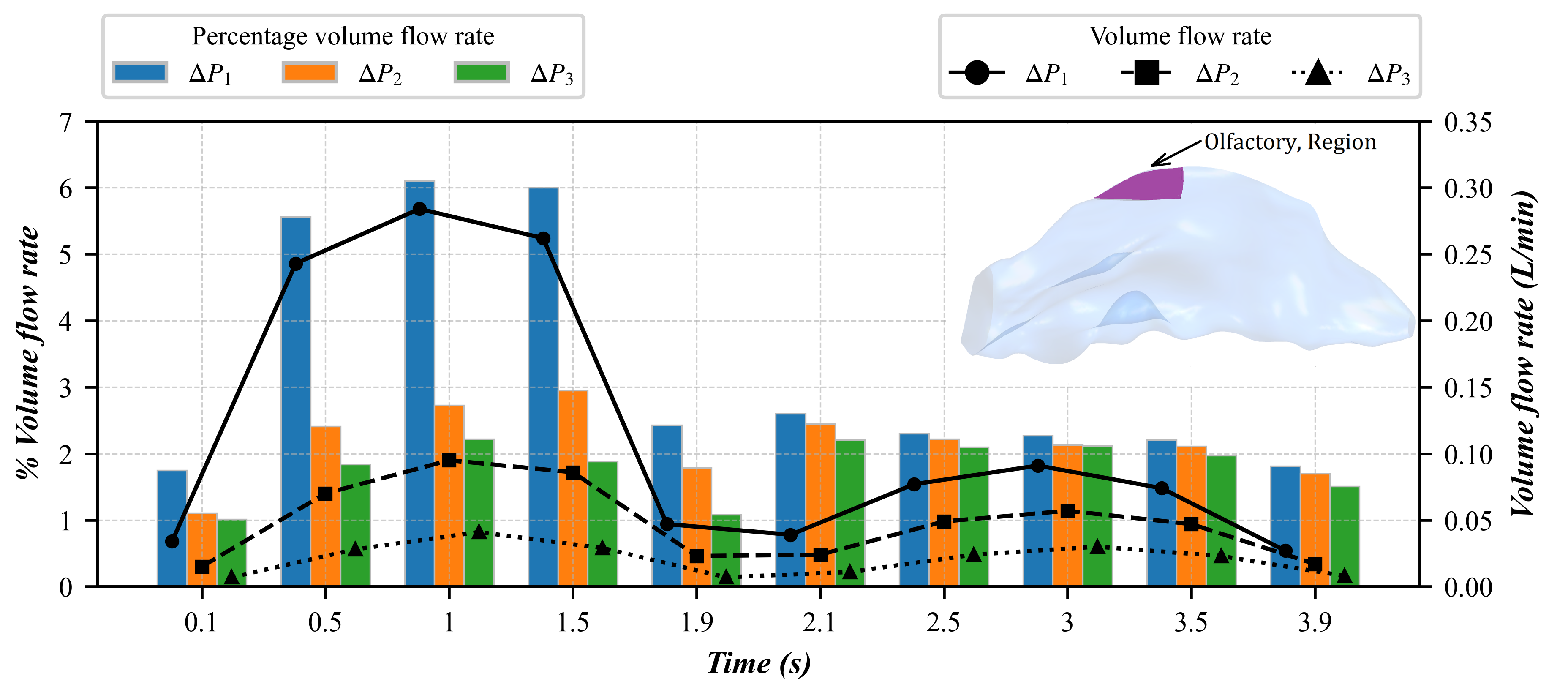}
    \caption{Temporal variation of absolute and percentage airflow through the olfactory region under the three pressure-driven breathing conditions.}
	\label{fig:olfaction_flow}
\end{figure*}

The breathing-phase asymmetry also influences airflow transport to the olfactory region. Figure~\ref{fig:olfaction_flow} shows the absolute olfactory flow rate and its percentage of the total nasal flow during the breathing cycle. During inhalation, increasing breathing intensity substantially enhances olfactory ventilation, with the high-pressure condition reaching approximately 6\% of the total nasal flow near peak inhalation. This increase is consistent with the stronger penetration of inspiratory flow into the superior nasal cavity observed earlier (Fig.~\ref{fig:phase_airflow}). In contrast, during exhalation the olfactory fraction remains close to 2--3\% and varies comparatively little with breathing intensity. The observed increase in olfactory-region flow with breathing intensity is consistent with~\citet{ruttgers2025comparative}, who reported increased upper-region airflow at elevated Reynolds numbers. Thus, the inhalation--exhalation asymmetry produced by pressure-driven breathing extends beyond the global flow rate and resistance, and directly influences regional airflow delivery to the olfactory region.

\section{Discussion} \label{sec:Discussion}

The principal finding of the present study is that pressure-driven breathing produces an inherent asymmetry between inhalation and exhalation \cite{horschler2010assumption,han2023investigation,shim2023numerical} in a population-derived median unilateral nasal cavity. Under identical imposed pressure magnitudes, inhalation generates a larger flow rate and lower resistance than exhalation \cite{Viani_Jones_Clarke_1990}, demonstrating that the pressure--flow response depends on flow direction and nasal geometry. Previous nasal CFD studies have shown that airway geometry strongly influences pressure loss, flow distribution, and resistance \citep{doorly2008mechanics,zhao2014normal,li2017computational}, while transient studies have demonstrated phase-dependent variations that are not represented by steady formulations \citep{lee2010unsteady,moniripiri2021numerical}.
The present results extend these observations by showing that, when pressure rather than flow rate is prescribed, the inhalation--exhalation difference in flow rate emerges naturally from the aerodynamic response of the nasal passage. This distinction is suppressed when identical inhalation and exhalation flow rates are imposed as boundary conditions.

The predicted unilateral resistance ($0.57$--$1.22~\mathrm{Pa{\cdot}s/mL}$) lies within the experimentally reported healthy unilateral range \citep{eccles1996changes}, supporting the physical consistency of the pressure-driven formulation. The accompanying WSS and pressure-loss distributions further indicate that the greater expiratory resistance cannot be explained by wall viscous loading alone, emphasizing the role of direction-dependent interaction between airflow and nasal geometry. The same phase asymmetry also has a regional consequence: increasing inspiratory intensity increases airflow delivery to the superior and olfactory regions, with approximately 6\% of the total flow reaching the olfactory region under the high-pressure condition. This is consistent with previous observations that olfactory function depends on effective airflow access to the olfactory epithelium \citep{sobel1998sniffing}. Thus, pressure-driven transient modelling provides information not only on global nasal resistance but also on the phase-dependent regional redistribution of airflow.

\section{Conclusions}\label{sec4}

Time-resolved CFD was used to investigate pressure-driven airflow through a representative unilateral human nasal cavity over a complete breathing cycle. The pressure-driven formulation revealed a distinct asymmetry between inhalation and exhalation: for the same imposed pressure magnitude, inhalation produced a larger flow rate and lower aerodynamic resistance than exhalation. This direction-dependent response was accompanied by differences in airflow organization, Reynolds number, and wall shear stress, demonstrating the important role of breathing phase in unilateral nasal aerodynamics.

Nasal resistance increased with breathing intensity, while the opposing breathing-phase trends in wall shear stress and pressure distribution indicated that the resistance asymmetry cannot be explained by wall shear alone. Breathing phase also influenced regional airflow distribution. Stronger inspiratory flow penetrated further into the superior nasal cavity, and the fraction of total airflow reaching the olfactory region increased to approximately 6\% under the high-pressure condition, whereas the corresponding expiratory fraction remained comparatively insensitive to breathing intensity. These findings demonstrate that pressure-driven transient modelling can capture geometry-dependent differences between inhalation and exhalation that are suppressed when identical flow rates are prescribed for the two breathing phases.

Although the standardized geometry is population-derived rather than patient-specific, it represents a single median anatomical configuration and therefore does not capture intersubject variability in nasal morphology or quantify the population distribution of the predicted aerodynamic response. The analysis is additionally restricted to a unilateral passage with rigid walls and a laminar airflow formulation; bilateral flow partitioning, mucosal compliance, and possible localized transitional effects were not considered.
Equal inhalation and exhalation durations were prescribed to isolate the aerodynamic response, whereas physiological breathing commonly exhibits unequal phase durations. Future work should extend the pressure-driven framework to subject-specific bilateral geometries and more physiological breathing waveforms, with higher-fidelity flow modelling where required.

\begin{acknowledgments}
MS thanks the Ministry of Education, Government of India, for scholarship support during his PhD. The authors acknowledge the National Supercomputing Mission for computing resources on `PARAM Himalaya' cluster at IIT Mandi, supported by C-DAC, MeiTY, and DST, Government of India.
\end{acknowledgments}


\section*{Conflict of Interest}
The authors have no conflicts to disclose.

\section*{Author Contributions}
\noindent
\textbf{Makesh Singh:} Conceptualization, Formal analysis, Investigation, Methodology, Validation, Visualization, Writing -- original draft.

\noindent
\textbf{Shashwat Bhattacharya:} Resources, Writing -- review \& editing, Supervision.

\noindent
\textbf{Gaurav Bhutani:} Formal analysis, Resources, Writing -- review \& editing, Supervision.


\section*{Data Availability Statement}
The data that support the findings of this study are available from the corresponding author upon reasonable request.


\vspace{1cm}
\noindent
\textbf{REFERENCES}\\
\bibliography{aipsamp}

@article{subramaniam1998computational,
  title={Computational fluid dynamics simulations of inspiratory airflow in the human nose and nasopharynx},
  author={Subramaniam, Ravi P and Richardson, Regina B and Morgan, Kevin T and Kimbell, Julia S and Guilmette, Raymond A},
  journal={Inhal. Toxicol.},
  volume={10},
  number={2},
  pages={91--120},
  year={1998},
  doi={10.1080/089583798197772},
  publisher={Taylor \& Francis}
}

@article{eccles2000,
  title={Nasal airflow in health and disease},
  author={Eccles, Ronald},
  journal={Acta Otolaryngol.},
  volume={120},
  number={5},
  pages={580--595},
  year={2000},
  doi={10.1080/000164800750000388},
  publisher={Taylor \& Francis}
}

@article{doorly2008mechanics,
  title={Mechanics of airflow in the human nasal airways},
  author={Doorly, Denis J and Taylor, Donal Joseph and Schroter, Robert C},
  journal={Respir. Physiol. Neurobiol.},
  volume={163},
  number={1-3},
  pages={100--110},
  year={2008},
  doi={10.1016/j.resp.2008.07.027},
  publisher={Elsevier}
}

@article{zhao2014normal,
  title={What is normal nasal airflow? A computational study of 22 healthy adults},
  author={Zhao, Kai and Jiang, Jianbo},
  journal={Int. Forum Allergy Rhinol.},
  volume={4},
  pages={435--446},
  year={2014},
  doi={10.1002/alr.21319},
  organization={Wiley Online Library}
}

@article{kim2013patient,
  title={Patient specific CFD models of nasal airflow: overview of methods and challenges},
  author={Kim, Sung Kyun and Na, Yang and Kim, Jee-In and Chung, Seung-Kyu},
  journal={J. Biomech.},
  volume={46},
  number={2},
  pages={299--306},
  year={2013},
  doi={10.1016/j.jbiomech.2012.11.022},
  publisher={Elsevier}
}

@article{keustermans2018high,
  title={High quality statistical shape modelling of the human nasal cavity and applications},
  author={Keustermans, William and Huysmans, Toon and Danckaers, Femke and Zarowski, Andrzej and Schmelzer, Bert and Sijbers, Jan and Dirckx, Joris JJ},
  journal={R. Soc. Open Sci.},
  volume={5},
  number={12},
  pages={181558},
  year={2018},
  doi={10.1098/rsos.181558},
  publisher={The Royal Society}
}

@article{basu2018computational,
  title={On computational fluid dynamics models for sinonasal drug transport: Relevance of nozzle subtraction and nasal vestibular dilation},
  author={Basu, Saikat and Frank-Ito, Dennis O and Kimbell, Julia S},
  journal={Int. J. Numer. Methods Biomed. Eng.},
  volume={34},
  number={4},
  pages={e2946},
  year={2018},
  doi={10.1002/cnm.2946},
  publisher={Wiley Online Library}
}

@article{asama2022analysis,
  title={Analysis of conductive olfactory dysfunction using computational fluid dynamics},
  author={Asama, Youji and Furutani, Akiko and Fujioka, Masato and Ozawa, Hiroyuki and Takei, Satoshi and Shibata, Shigenobu and Ogawa, Kaoru},
  journal={PLoS One},
  volume={17},
  number={1},
  pages={e0262579},
  year={2022},
  doi={10.1371/journal.pone.0262579},
  publisher={Public Library of Science San Francisco, CA USA}
}

@article{lee2010unsteady,
  title={Unsteady flow characteristics through a human nasal airway},
  author={Lee, Jong-Hoon and Na, Yang and Kim, Sung-Kyun and Chung, Seung-Kyu},
  journal={Respir. Physiol. Neurobiol.},
  volume={172},
  number={3},
  pages={136--146},
  year={2010},
  doi={10.1016/j.resp.2010.05.010},
  publisher={Elsevier}
}

@article{gu2019,
  title={Numerical investigation of unsteady particle deposition in a realistic human nasal cavity during inhalation},
  author={Gu, Xin and Wen, Jian and Wang, Mengmeng and Jian, Guanping and Zheng, Guoxi and Wang, Simin},
  journal={Exp. Comput. Multiphase Flow},
  volume={1},
  number={1},
  pages={39--50},
  year={2019},
  doi={10.1007/s42757-019-0007-0},
  publisher={Springer}
}

@article{moniripiri2021numerical,
  title={Numerical simulation of unsteady airflow in a nasal cavity for various sizes of maxillary sinus opening in a virtual endoscopic surgery},
  author={Moniripiri, Mohammad and Amjadimanesh, Hossein and Faramarzi, Mohammad and Sadrizadeh, Sasan and Abouali, Omid},
  journal={Respir. Physiol. Neurobiol.},
  volume={291},
  pages={103690},
  year={2021},
  doi={10.1016/j.resp.2021.103690},
  publisher={Elsevier}
}

@inproceedings{shahed2024analysis,
  title={Analysis of Flow Field Under Unsteady Breathing Conditions in a Realistic Human Nasal Cavity},
  author={Shahed, Syeda Reham and Tian, Lin and Dong, Jingliang and Tu, Jiyuan and Zhang, Ya and Ma, Ruiping and Ahmadi, Goodarz},
  booktitle={Fluids Engineering Division Summer Meeting},
  volume={88124},
  pages={V001T04A014},
  year={2024},
  organization={American Society of Mechanical Engineers}
}

@article{patel2015simulating,
  title={Simulating the nasal cycle with computational fluid dynamics},
  author={Patel, Ruchin G and Garcia, Guilherme JM and Frank-Ito, Dennis O and Kimbell, Julia S and Rhee, John S},
  journal={Otolaryngol. Head Neck Surg.},
  volume={152},
  number={2},
  pages={353--360},
  year={2015},
  doi={10.1177/019459981455938},
  publisher={SAGE Publications Sage CA: Los Angeles, CA}
}

@article{frank2019computational,
  title={Computational analysis of the mature unilateral cleft lip nasal deformity on nasal patency},
  author={Frank-Ito, Dennis O and Carpenter, David J and Cheng, Tracy and Avashia, Yash J and Brown, David A and Glener, Adam and Allori, Alexander and Marcus, Jeffrey R},
  journal={Plast. Reconstr. Surg. Glob. Open},
  volume={7},
  number={5},
  pages={e2244},
  year={2019},
  doi={10.1097/GOX.0000000000002244},
  publisher={LWW}
}

@article{faramarzi2014numerical,
  title={Numerical investigation of the flow field in realistic nasal septal perforation geometry},
  author={Faramarzi, Mohammad and H Baradaranfar, Mohammad and Abouali, Omid and Atighechi, Saeid and Ahmadi, Goodarz and Farhadi, Pejman and Keshavarzian, Erfan and Behniafard, Nasim and Baradaranfar, Amin},
  journal={Allergy Rhinol.},
  volume={5},
  number={2},
  pages={ar--2014},
  year={2014},
  doi={10.2500/ar.2014.5.0090},
  publisher={SAGE Publications Sage CA: Los Angeles, CA}
}

@article{liu2009standardized,
  title={Creation of a standardized geometry of the human nasal cavity},
  author={Liu, Y. and Johnson, Matthew R. and Matida, Edward A. and Kherani, S. and Marsan, J.},
  journal={J. Appl. Physiol.},
  volume={106},
  number={3},
  pages={784--795},
  year={2009},
  doi={10.1152/japplphysiol.90376.2008},
  publisher={American Physiological Society}
}

@article{li2017computational,
  title={Computational modeling and validation of human nasal airflow under various breathing conditions},
  author={Li, Chengyu and Jiang, Jianbo and Dong, Haibo and Zhao, Kai},
  journal={J. Biomech.},
  volume={64},
  pages={59--68},
  year={2017},
  doi={10.1016/j.jbiomech.2017.08.031},
  publisher={Elsevier}
}

@article{schillaci2022importance,
  title={Importance of the numerical schemes in the CFD of the human nose},
  author={Schillaci, Andrea and Quadrio, Maurizio},
  journal={J. Biomech.},
  volume={138},
  pages={111100},
  year={2022},
  publisher={Elsevier}
}

@article{sullivan1991steady,
  title={Steady and oscillatory transnasal pressure-flow relationships in healthy adults},
  author={Sullivan, Kevin J and Chang, HK},
  journal={J. Appl. Physiol.},
  volume={71},
  number={3},
  pages={983--992},
  year={1991},
  doi={10.1152/jappl.1991.71.3.983}
}

@article{motamedi2021reference,
  title={Reference equations for tidal breathing parameters using structured light plethysmography},
  author={Motamedi-Fakhr, Shayan and Iles, Richard and Barker, Nicki and Alexander, John and Cooper, Brendan G},
  journal={ERJ Open Res.},
  volume={7},
  number={2},
  pages={00050--2021},
  year={2021},
  publisher={European Respiratory Society}
}

@article{eccles1996changes,
  title={Changes in the amplitude of the nasal cycle associated with symptoms of acute upper respiratory tract infection},
  author={Eccles, R and Reilly, M and Eccles, KSJ},
  journal={Acta Otolaryngol.},
  volume={116},
  number={1},
  pages={77--81},
  year={1996},
  doi={10.3109/00016489609137717},
  publisher={Taylor \& Francis}
}

@article{Viani_Jones_Clarke_1990,
  title={Nasal airflow in inspiration and expiration},
  volume={104},
  DOI={10.1017/S0022215100112915},
  number={6},
  journal={J. Laryngol. Otol.},
  author={Viani, Laura and Jones, Andrew S. and Clarke, Ray},
  year={1990},
  pages={473--476}
}

@article{sobel1998sniffing,
  title={Sniffing and smelling: separate subsystems in the human olfactory cortex},
  author={Sobel, Noam and Prabhakaran, Vivek and Desmond, John E and Glover, Gary H and Goode, RL and Sullivan, Edith V and Gabrieli, John DE},
  journal={Nature},
  volume={392},
  number={6673},
  pages={282--286},
  year={1998},
  publisher={Nature Publishing Group UK London}
}

@article{niegodajew2025flow,
  title={Flow patterns and vortex formation mechanisms inside a human nasal cavity},
  author={Niegodajew, Pawe{\l}},
  journal={Phys. Fluids},
  volume={37},
  number={2},
  year={2025},
  publisher={AIP Publishing}
}

@article{shim2023numerical,
  title={Numerical simulation of virus-laden aerosol transmission in real human respiratory airways},
  author={Shim, Gihun and Narayanan, Sai Ranjeet and Yang, Suo},
  journal={Phys. Fluids},
  volume={35},
  number={10},
  year={2023},
  publisher={AIP Publishing}
}

@article{tripathy2024effect,
  title={Effect of various septal deviations and corrections on nasal aero and particle dynamics: An in silico and in vitro investigation},
  author={Tripathy, Kartika Chandra and Bhandari, Ajay},
  journal={Phys. Fluids},
  volume={36},
  number={9},
  year={2024},
  publisher={AIP Publishing}
}

@article{han2023investigation,
  title={Investigation of the effects of nasal surgery on nasal cavity flow using magnetic resonance velocimetry and computational fluid dynamics},
  author={Han, Kyuho and Lee, Sung-Gwang and Kim, Kwanwoo and Jeong, Baren and Paek, Munyoung and Lee, Whal and Hwang, Wontae},
  journal={Phys. Fluids},
  volume={35},
  number={11},
  year={2023},
  publisher={AIP Publishing}
}

@article{van2021pressure,
  title={Pressure distribution and flow dynamics in a nasal airway using a scale resolving simulation},
  author={Van Strien, James and Shrestha, Kendra and Gabriel, Sargon and Lappas, Petros and Fletcher, David F and Singh, Narinder and Inthavong, Kiao},
  journal={Phys. Fluids},
  volume={33},
  number={1},
  year={2021},
  publisher={AIP Publishing}
}

@article{ruttgers2025comparative,
  title={Comparative analysis of the flow in a realistic human airway},
  author={R{\"u}ttgers, Mario and Vorspohl, Julian and Mayolle, Luca and Johanning-Meiners, Benedikt and Krug, Dominik and Klaas, Michael and Meinke, Matthias and Lee, Sangseung and Schr{\"o}der, Wolfgang and Lintermann, Andreas},
  journal={Phys. Fluids},
  volume={37},
  number={12},
  year={2025},
  publisher={AIP Publishing}
}

@article{horschler2003numerical,
  title={Numerical simulation of the flow field in a model of the nasal cavity},
  author={H{\"o}rschler, Ingolf and Meinke, Matthias and Schr{\"o}der, Wolfgang},
  journal={Comput. Fluids},
  volume={32},
  number={1},
  pages={39--45},
  year={2003},
  publisher={Elsevier}
}

@article{horschler2010assumption,
  title={On the assumption of steadiness of nasal cavity flow},
  author={H{\"o}rschler, Ingolf and Schr{\"o}der, W and Meinke, M},
  journal={J. Biomech.},
  volume={43},
  number={6},
  pages={1081--1085},
  year={2010},
  publisher={Elsevier}
}

\end{document}